\documentclass[review]{elsarticle}
\usepackage{graphicx}
\usepackage{epsfig,graphics,subfigure,psfrag,amsmath,amssymb}
\usepackage{lineno,hyperref}
\modulolinenumbers[5]

\begin{document}

\begin{frontmatter}

\title{Wheeler-DeWitt equation rejects quantum effects of grown-up universes as a candidate for dark energy}

\author{Dongshan He}
\address{College of Physics \& Electronic Engineering, Xianyang Normal University. Xianyang 712000, China}

\author{Qing-yu Cai\corref{mycorrespondingauthor}}
\cortext[mycorrespondingauthor]{Corresponding author}
\ead{qycai@wipm.ac.cn}
\address{State Key Laboratory of Magnetic Resonances and Atomic and Molecular Physics,
Innovation Academy for Precision Measurement Science and Technology, Chinese Academy of Sciences, Wuhan 430071, China \\
Peng Huanwu Center for Fundamental Theory, Hefei, Anhui 230026, China}


%
%
%

\begin{abstract}
In this paper, we study the changes of quantum effects of a growing universe by using
Wheeler-DeWitt equation (WDWE) together with de Broglie-Bohm quantum trajectory approach.
From WDWE, we obtain the quantum modified Friedmann equations which have additional
terms called quantum potential compared to standard Friedmann equations.
The quantum potential governs the behavior of the early universe, providing energy
for inflation, while it decreases rapidly as the universe grows.
The quantum potential of the grown-up universe is much smaller than that required for
accelerating expansion. This indicates that quantum effects of our universe cannot
be treated as a candidate for dark energy.
\end{abstract}


\end{frontmatter}


\section{Introduction}

It has been widely accepted that our universe has been expanding at accelerating
rates in recent epochs. Dark energy was suggested to explain the acceleration of
the universe. Although scientists are still plagued by its intangibility,
the quantum effect of the universe is considered as the most likely candidate for
dark energy~\cite{pr03}.
In the 1980s, quantum mechanics had already been applied to study quantum effects
of the universe, for example, the formation of the universe and its early evolution~\cite{hh83,swh84,av86}.
In quantum cosmology, the universe is described by the universe wavefunction rather
than classical spacetime. The wavefunction of the universe that contains all the
information of the universe should satisfy the quantum gravity equation, or called
Wheeler-DeWitt equation (WDWE)~\cite{bd67}. Due to the ADM decomposition, the WDWE
equation is time-independent, which prevents us from studying the quantum effects
as well as dynamics of the universe directly. In order to find out how quantum effects
change as the universe grows, de Broglie-Bohm quantum trajectory approach can be used
to evolve WDWE.

In de Broglie-Bohm quantum trajectory theory~\cite{bd52,prh93}, the quantum potential
represents the quantum effects of a quantum system. Similarly, the quantum potential
of the universe describes the quantum effects of the universe. Previous studies of the
quantum-to-classical transition show that it is the quantum potential that dominates
the quantum effects of a physical system and trends toward classical behaviors when the
quantum potential becomes negligible~\cite{lcz09}.
By solving the quantum potential of the universe, the values of the
quantum effects at different stages of the universe can be obtained.
It can then be determined that whether quantum effects can provide
enough power for the accelerating expansion of the universe or not.

In this article, with de Broglie-Bohm quantum trajectory theory, the quantum
modified Friedmann equations are derived from the WDWE.
Then the changes of the quantum potential at different stage of the universe are studied,
and three typical cosmological models, including the radiation-dominated universe, the
vacuum universe and the universe with a scalar-field, are considered.
Finally, we discuss and conclude.

\section{WDWE of the universe}\label{section:WDWE}

Assumed to be homogeneous and isotropic, the universe can be
described by a minisuperspace model~\cite{npn13,npn12,apk97} with one parameter, the scale factor $a$.
The Einstein-Hilbert action for the model can be written as
\begin{equation}
S_{{\rm EH}}=\int\left(  \frac{\mathcal{R} c^{3}}{16\pi G}-\frac{\rho}{\sigma^2 c}\right)
\sqrt{-g}d^{4}x, \label{action}
\end{equation}
where $\rho$ represents the energy density of the universe, and
the constant before $\rho$ is chosen for the sake of convenience. Since
the universe is homogeneous and isotropic, the metric of the universe in the
minisuperspace model is given by
\begin{equation}
ds^{2}=\sigma^{2}\left[  -N^{2}(t)c^{2}dt^{2}+a^{2}(t)d\Omega_{3}^{2}\right]
. \label{metric}%
\end{equation}
Here, $d\Omega_{3}^{2}=dr^{2}/(1-kr^{2})+r^{2}(d\theta^{2}+\sin^{2}\theta
d\phi^{2})$ is the metric on a unit three-sphere, $N(t)$ is an arbitrary lapse
function, and $\sigma^{2}=2/3\pi$ is just a normalization factor for simplifying
the formula~\cite{av88}. Note that $r$ is dimensionless
and the scale factor $a(t)$ has dimensions of length~\cite{sw72}.
From Eq.~(\ref{metric}), one can obtain the scalar curvature
\begin{equation}
\mathcal{R}=6\frac{\ddot{a}}{\sigma^{2}c^{2}N^{2}a}+6\frac{\dot{a}^{2}}{\sigma^{2}
c^{2}N^{2}a^{2}}+\frac{6k}{\sigma^{2}a^{2}}, \label{Ricci}%
\end{equation}
where the dot denotes the derivative with respect to the time, $t$.
Inserting Eqs.~(\ref{metric}) and~(\ref{Ricci}) into Eq.~(\ref{action}), one obtains
\begin{eqnarray*}
S_{EH}  &=& \int\frac{6\sigma^{2}Nc^{4}}{16\pi G}\left(  \frac{a^{2}\ddot{a}}%
{N^{2}c^{2}}+\frac{a\dot{a}^{2}}{N^{2}c^{2}}+ka-\frac{8\pi G\rho a^{3}%
}{3c^{4}}\right)  d^{4}x,\\
&=& \frac{Nc^{4}}{2G}\int\left(  -\frac{a\dot{a}^{2}}{N^{2}c^{2}}%
+ka-\frac{8\pi G\rho a^{3}}{3c^{4}}\right)  dt.
\end{eqnarray*}
Therefore, the Lagrangian of the universe can be written as
\begin{equation}
\mathcal{L}=\frac{Nc^{4}}{2G}\left(  ka-\frac{a\dot{a}^{2}}{N^{2}c^{2}}%
-\frac{8\pi G\rho a^{3}}{3c^{4}}\right) , \label{Lagrangian}%
\end{equation}
and the momentum $p_{a}$ can be obtained as
\begin{equation}
  p_{a}=\frac{\partial\mathcal{L}}{\partial\dot{a}}=-\frac{c^{2}a\dot{a}}{NG}.
\end{equation}
Taking $N=1$, the Hamiltonian is found to be
\begin{eqnarray}
\mathcal{H}&=&p_{a}\dot{a}-\mathcal{L} \nonumber\\
&=&-\frac{1}{2}\left(  \frac{Gp_{a}^{2}}{c^{2}a}+\frac{c^{4}ka}
{G}-\frac{8\pi \rho a^{3}}{3}\right) .\nonumber
\end{eqnarray}
With $\mathcal{H}\Psi=0$ and $p_{a}^{2}=-\hbar^{2}a^{-p}\frac{\partial}{\partial
a}(a^{p}\frac{\partial}{\partial a})$, one gets the WDW equation~\cite{bd67,swh84,av94},
\begin{equation}
\left(  \frac{\hbar^{2}}{m_{p}}\frac{1}{a^{p}}\frac{\partial}{\partial a}%
a^{p}\frac{\partial}{\partial a}-\frac{E_{p}}{l_{p}^{2}}ka^{2}+\frac
{8\pi \rho a^{4}}{3 l_{p}}\right)  \psi(a)=0. \label{wdwe1}%
\end{equation}
Here, $k=1,0,-1$ are for spatially closed, flat and open universe,
respectively. The factor $p$ represents the uncertainty in the choice of
operator ordering, $m_{p}$, $E_{p}$, $l_{p}$, and $t_{p}$ are the Planck mass,
Planck energy, Planck length, and Planck time, respectively.

\section{Quantum Modified Friedmann Equations}\label{section:quantum:friedmann:equation}

The complex function $\psi(a)$ can be rewritten as
\begin{equation}
\psi(a)=R(a)\exp[iS(a)/\hbar], \label{psi}%
\end{equation}
where $R$ and $S$ are real functions. Inserting $\psi(a)$
into Eq.~(\ref{wdwe1}) and separating the equation into real and imaginary
parts, one can obtain two equations~\cite{bd52,prh93}
\begin{eqnarray}
S^{\prime\prime}+2\frac{R^{\prime}S^{\prime}}{R}+\frac{p}{a}S^{\prime}  &=&0
,\label{wdwe2}\\
\frac{(S^{\prime})^{2}}{m_p}+U+Q  &=&0. \label{wdwe3}%
\end{eqnarray}
Here, the prime denotes derivatives with respect to $a$,
and $U(a)$ and $Q(a)$ are the classical potential and the quantum potential of the universe, respectively,
\begin{eqnarray}
U(a)&=&\frac{E_p k a^{2}}{l_p^2}-\frac{8\pi \rho a^4}{3l_p},\label{classical:potential}\\
Q(a)&=&-\frac{\hbar^2}{m_p}(\frac{R^{\prime\prime}}{R}+\frac{p}{a}\frac{R^{\prime}}{R}). \label{quantum:potential}%
\end{eqnarray}
Using Eq.~(\ref{wdwe2}), we can obtain another form of the quantum potential~\cite{hgc14},
\begin{equation}
Q(a)=-\frac{\hbar^2}{m_p}\left[\frac{-p^{2}+2p}{4a^{2}}+\frac{3(S^{\prime\prime})^{2}}{4(S^{\prime
})^{2}}-\frac{S^{\prime\prime\prime}}{2S^{\prime}}\right]. \label{quantum:potential2}%
\end{equation}
If one knows $R$ or $S$ of the wavefunction of the universe, then one can obtain the quantum potential
by using Eq.~(\ref{quantum:potential}) or Eq.~(\ref{quantum:potential2}).

By analogy with cases in non-relativistic particle physics and quantum field
theory in flat space-time, quantum trajectories can be obtained from the
guidance relation~\cite{npn13,lpg93},
\begin{eqnarray}
\frac{\partial\mathcal{L}}{\partial \dot{a}}  &=&-\frac{c^2}{G}a\dot{a}
=S^{\prime},\label{gr}\\
\dot{a}  &=&-\frac{G S^{\prime}}{c^2 a}. \label{gr2}%
\end{eqnarray}
From Eqs.~(\ref{wdwe3}) and~(\ref{gr2}), the Hubble parameter can be written as
\begin{equation}
H^2(t)=\left({\frac{\dot{a}}{a}}\right)^2=-\frac{G^2 m_{p}}{c^{4}} \frac{Q(a)+U(a)}
{a^{4}}. \label{Hubble}%
\end{equation}
Inserting Eq.~(\ref{classical:potential}) into Eq.~(\ref{Hubble}), we obtain the
first Friedmann equation except for one more term that contains the quantum potential $Q(a)$,
\begin{equation}\label{Friedmann1}
H^2=\frac{8\pi G \rho(a)}{3c^2}-\frac{kc^2}{a^2}-\frac{l_p^2}{m_p}\frac{Q(a)}{a^4}.
\end{equation}
By using the state equation $P=w \rho$ and energy conservation law $\rho(a) \propto a^{-3(w+1)}$,
the second quantum modified Friedmann equation (QMFE) can be attained as
\begin{equation}\label{Friedmann2}
\frac{\ddot{a}}{a}=-\frac{4\pi G}{3c^2}\left[\rho(a)+3P(a) \right]+\frac{l_p^2}{m_p}
\left[\frac{Q}{a^4}-\frac{Q'(a)}{2a^3}\right].
\end{equation}
These two equations above turn to classical Friedmann equation when all terms about
the quantum potential trend to zero, i.e., $Q(a)\rightarrow 0$ and $Q'(a)\rightarrow 0$.

\section{The changes of quantum potentials for growing universes}

In general, there are different models for the universe. Here we consider three
representative models, the universe with classical potential, the universe of vacuum, and
the universe with scalar field. We will study the quantum potential and the evolution of the
universe with these three different models, respectively.

\subsection{Quantum potential and the evolution of the universe with classical potential}

Let us study the evolution laws of the universe with the QMFE. For simplicity, we only
consider the case of the flat universe $k=0$ \footnote{For $k=\pm 1$, we can obtain similar results~\cite{hgc14}.}.
The WDWE for the flat universe with energy density $\rho_{n}$ can be written as
\begin{equation}
\left(  \frac{\hbar^{2}}{m_{p}}\frac{1}{a^{p}}\frac{\partial}{\partial a}%
a^{p}\frac{\partial}{\partial a}+\frac{8\pi \rho_{n}(a) a^{4}}{3 l_{p}}\right)  \psi(a)=0. \label{WDWEn}%
\end{equation}
Here $\rho_{n}(a)=3 \lambda_{n}E_p l_p^{n-3}/8\pi a^{n}$, and $\lambda_{n}$ is a
dimensionless parameter that relates to the energy density at $a=l_p$, with $n=4,3,0$
representing the universe dominated by radiation, matter and dark energy, respectively~\cite{dhc05},
and the explicit form of the corresponding classical potential is $U(a)=-\lambda_n E_p(a/l_p)^{4-n}$.
In principle, the universe contains all types of energy at the same stage; thus the energy
density should take the form of $\rho_=\rho_{0}+\rho_{3}+\rho_{4}$. In practice, the universe
is mainly dominated by one type of energy $\rho_n$ during a specific stage. During the expansion of
the universe, $n$ changes slowly from $n=4$ to $n=0$. For simplicity, we will take $\hbar=G=c=1$
in the following.

\begin{figure}[!hbt]
\includegraphics[width=8cm]{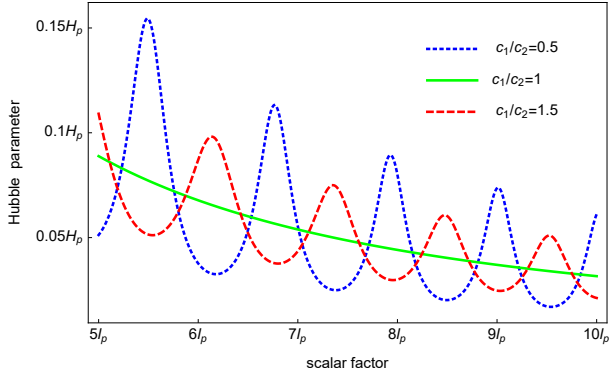}
\caption{(Color online)\, The change of Hubble parameters with different $c_1 / c_2$.
Here we have set $n=3$ and $p=-2$. More detailed calculations show
that the Hubble parameters oscillate violently when $c_1 / c_2\neq 1$,
regardless of the values of $n$ and $p$.} \label{fig:cHubble}
\end{figure}

For an arbitrary $n$, we can obtain general solutions of Eq.~(\ref{WDWEn}),
\begin{eqnarray}
\psi_{n}(a) =a^{\frac{1-p}{2}} \left[ {i c_{1}J_{\nu}\left(
\frac{\sqrt{\lambda_{n}}a^{3-n/2}}{3-n/2}\right)}\right. 
\left.{ +c_{2}Y_{\nu}
\left( \frac{\sqrt{\lambda_{n}}a^{3-n/2}}{3-n/2}\right)}\right].
\label{wave:function}
\end{eqnarray}
Here, $J_{\alpha}(x)'s$ are Bessel functions of the first kind, $Y_{\alpha}(x)'s$
are Bessel functions of the second kind, and $\nu=\left\vert(1-p)/(n-6)\right\vert$.
It should be noted that the wave function (\ref{wave:function}) diverges at $a\rightarrow0$,
when $p=1$~\footnote{The divergence of the wavefunction in Eq.~(\ref{wave:function}) is due to the divergence of Neumann functions $Y_{\nu}$.
It is clear that the divergence doesn't appear when $c_2=0$. But, as will be discussed in Sec.~\ref{Discussion:and:conclusion},
there is no acceleration for the universe with a purely imaginary (or purely real) wavefunction.}.
$c_{1}$ and $c_{2}$ are constants that should be determined by boundary conditions.
When $c_1\neq c_2$, Hubble parameters will oscillate with the expansion of the universe,
and the oscillation frequencies will increase as the universe growing up, which seems quite
unreasonable. Numerical solutions are shown in FIG.~\ref{fig:cHubble}.
In the following, we set $c_{1}=c_{2}$ for reasonable solutions.

First, we study the evolution of the early universe ($a\ll l_p$) by using QMFE Eqs.~(\ref{Friedmann1})
and ~(\ref{Friedmann2}). From the wavefunction Eq.~(\ref{wave:function}), we can obtain
\begin{equation}\label{R:small}
R=a^{\frac{1-p}{2}}\sqrt{J^2_{\nu}\left(
\frac{\sqrt{\lambda_{n}} a^{3-n/2}}{3-n/2}\right)+Y^2_{\nu}
\left( \frac{\sqrt{\lambda_{n}} a^{3-n/2}}{3-n/2}\right)}.
\end{equation}
Inserting Eq.~(\ref{R:small}) into Eq.~(\ref{quantum:potential}), the
quantum potential of the universe can be obtained as
\begin{equation}
Q(a)= \frac{ \lambda \csc^4(\pi \nu ) \left[J^2_{-\nu }(\sqrt{\lambda_4}a)+J^2_{\nu }(\sqrt{\lambda_4}a)-2\cos(\pi \nu )J_{\nu }(\sqrt{\lambda_4}a) J_{-\nu }(\sqrt{\lambda_4}a)\right]^2-4\pi^{-2}a^{-2}}
{\left[J^2_{\nu }(\sqrt{\lambda_4}a)+Y^2_{\nu }(\sqrt{\lambda_4}a)\right]^2}.  \nonumber
\end{equation}
Expanding $Q(a)$ in series at $a=0$ and keeping it to the lowest order of $a$, one can get the form of quantum potential
of the early universe as
\begin{equation}\label{Q:small}
Q(a)=\left\{\begin{aligned}\lambda_{4}-\frac{\pi^2 \lambda_{4}^{2\nu}}{\Gamma^4(\nu)} \left(\frac{a}{2}\right)^{4\nu-2}, \ \textit{for} \ a\ll l_p \textit{,\ }p\neq 1. \\
-\frac{\pi^{2}}{4a^2 \ln^4 a},\,\,\,\,\,\,\,\,\,\,\,\,\,\,\,\,\,\,\,\,\,\,\,\,\,\,\,\,\, \ \textit{for} \ a\ll l_p \textit{,\ }p=1.
\end{aligned}
\right.
\end{equation}
Here, we have considered the case of $n=4$, since the classical potential of the universe is dominated by
the radiation at the very early stage, and $\lambda_4$ equals the classical potential
$\lambda_4=U(a)=-8\pi \rho_4 (a)a^4/3=Constant$. The ordering factor $p=-2$ can be determined by the
boundary conditions of the universe~\cite{hgc15}. Inserting the quantum potential in Eq.~(\ref{Q:small})
for the specific case of $p=-2$ into the first QMFE Eq.~(\ref{Friedmann1}), we obtain
\begin{eqnarray}\label{H:small}
H^2&=\frac{8\pi \rho_4 (a)}{3}-\frac{Q(a)}{a^4}, \\
&=\lambda_{4}^3=Constant.
\end{eqnarray}
Therefore, the universe will expand exponentially during the early stage:
\begin{equation}\label{a:small}
a(t)=e^{H(t+t_0)}.
\end{equation}
By the calculations above, we have proven that the quantum potential plays a crucial role in
the early universe to promote the expansion of the universe. This exponential expansion solution
is consistent with quantum trajectory theory and the dynamical interpretation of the universe
wavefunction~\cite{hgc14,hgc15,hc15}.

\begin{figure}[!hbt]
\includegraphics[width=8cm]{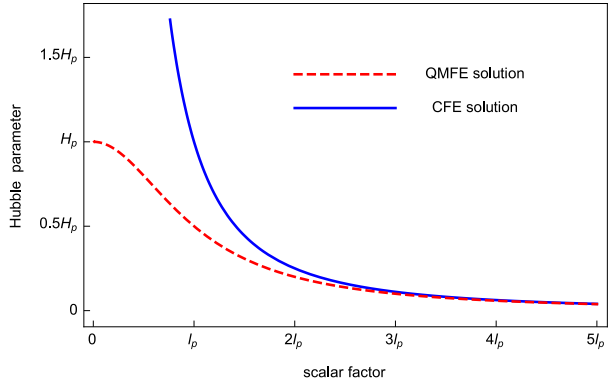}
\caption{(Color online)The evolution of Hubble parameter $H$ for the quantum modified Friedmann
equation (QMFE) solution and classical Friedmann equation (CFE) solution.} \label{fig:Hubble:small}
\end{figure}

\begin{figure}[!hbt]
\includegraphics[width=8cm]{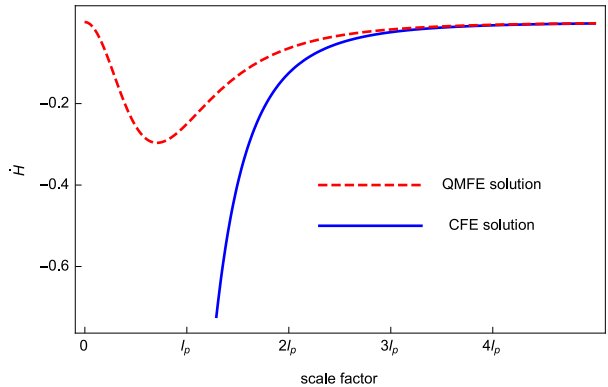}
\caption{(Color online)The evolution of $\dot{H}$ for the quantum modified Friedmann equation (QMFE)
solution and classical Friedmann equation (CFE) solution. }\label{fig:Hubble:dot}
\end{figure}

Next, we consider the wavefunction for the grown-up universe, i.e., $a\gg l_p$.
For $x\gg\left\vert \nu^{2}-1/4\right\vert $, the Bessel functions take the following asymptotic forms:
\begin{eqnarray*}
J_{\nu}(x)  &  \sim & \sqrt{\frac{2}{\pi x}}\cos(x-\nu\pi/2-\pi/4),\\
Y_{\nu}(x)  &  \sim & \sqrt{\frac{2}{\pi x}}\sin(x-\nu\pi/2-\pi/4).
\end{eqnarray*}
If the parameters $c_1$ and $c_2$ in Eq.~(\ref{wave:function})
take the value $c_{1}=c_{2}=c_{0}\sqrt{\pi/2}$, the wavefunction can
be rewritten as
\begin{equation}
\psi_{n}(a)=c_{0}a^{\frac{n-2p-4}{4}}\exp\left[  \frac{-i\sqrt{\lambda_{n}%
}a^{3-n/2}}{3-n/2}+i\theta\right], \label{wave:function:large}
\end{equation}
where $\theta=(3n-2p-16)\pi/(4n-24)$, and  $S^{\prime}<0$.
From Eq.~(\ref{gr2}), we know that the wavefunction in Eq.~(\ref{wave:function:large})
describes an expanding universe as suggested by Vilenkin~\cite{av88}.
With the wavefunction above, we can obtain
\begin{equation}
R_n(a)=c_{0}a^{\frac{n-2p-4}{4}}
\end{equation}
Together with Eq.~(\ref{quantum:potential}), the above equation gives
\begin{equation}\label{quantum:potential:large}
Q(a)=\frac{4 (p-1)^2 - (n-6)^2}{16 a^2}, \ \textit{for} \ a\gg l_p.
\end{equation}
It is obvious that $ Q/a^4 \ll \rho(a)$ and $Q'/a^3 \ll \rho(a), p(a)$ for the large universe.
Therefore, we conclude that the QMFE gradually turns into the classical Friedmann equations
for $a\gg l_p$; the evolution laws of the universe obtained from QMFE are completely
consistent with the solution of the classical Friedmann equations when $a\gg l_p$.

Numerical solutions of the classical and quantum modified Friedmann equation are shown
in Fig.~\ref{fig:Hubble:small} and Fig.~\ref{fig:Hubble:dot}. From the solution of the
classical Friedmann equation, one finds that $H$ and $\dot{H}$ are infinities for $a\rightarrow 0$,
which lead to a big-bang singularity. It is interesting that $H$ and $\dot{H}$ are finite when
the QMFEs are considered. In this way, we have shown that QMFE
can eliminate the big-bang singularity, in principle.

\subsection{Quantum potential and the evolution of the vacuum universe}

It is widely believed that the total energy of the universe is zero\cite{pcd84}, and the
universe may be a vacuum fluctuation~\cite{et73}. The wavefunction of the vacuum universe
satisfies Eq.~(\ref{wdwe1}) for $\rho=0$. For simplicity, we consider the flat universe
$k=0$ \footnote{For $k=\pm 1$ the wavefunction of the vacuum universe is similar to
the solution Eq.~(\ref{wave:function}) with $n=2$\cite{hgc14}.}. In this case, the analytic
solution of Eq. (\ref{wdwe1}) is
\begin{equation}
\psi(a)=ic_{1}\frac{a^{1-p}}{1-p}-c_{2},\ \textit{for}\ \  p\neq 1. \label{psi0}%
\end{equation}
With the wavefunction above, we can obtain
\begin{eqnarray} \label{vacuum rs}
S&=\tan^{-1}[-\frac{c_{1}}{c_{2}}\frac{a^{1-p}}{1-p}], \,\,\,\, p\neq 1, \\
R&=\sqrt{c_{2}^{2}+(c_{1}\frac{a^{1-p}}{1-p})^{2}}, \,\,\,\, p\neq 1.  \label{vacuum rs}
\end{eqnarray}
Using the guidance relation (\ref{gr2}), we can get the general form of the Bohmian
trajectories as
\begin{equation}\nonumber
a(t)=\left\{\begin{aligned}\left[\frac{c_1}{c_2}(3-|1-p|)(t+t_0)\right]^{\frac{1}{3-|1-p|}}, \, |1-p|\neq 0,3, \nonumber \\
e^{c_1(t+t_0)/c_2},\,\,\,\,\,\,\,\,\,\,\,\,\,\,\,\,\,\,\,\,\,\,\,\,\,\,\,\,
\,\,\,\,\,\,\,\,\,\,\,\,\,\,\,\,\,\,\,\,\,\,
|1-p|=3. \nonumber
\end{aligned}
\right.
\end{equation}
Inserting Eq.~(\ref{vacuum rs}) into Eq.~(\ref{quantum:potential}), the quantum potential
can be obtained as $$Q(a)=-\frac{c_1^2 c_2^2 (p-1)^4 a^{2 p}}{\left(c_2^2 (p-1)^2 a^{2 p}+a^2 c_1^2\right){}^2}.$$
For the small universe ($a\ll 1$), the quantum potential approximates to $Q(a)\sim -a^{-2+2|p-1|}$,
and for the large universe ($a\gg1$), the quantum potential is about $Q(a)\sim -a^{-2-2|p-1|}.$
Likewise, only the ordering factor takes the value $p=-2$ (or 4) and  $c_{1}/c_{2}>0$, will the small
true vacuum bubble expand exponentially.

For the case of exponential expansion, the quantum potential for the vacuum bubble can be obtained as $Q(a\rightarrow
0)=-(c_{1}/c_{2})^{2}a^{4}$, while the classical potential is $V(a)=0$. This
definitely indicates that quantum potential $Q(a)$ is the origin of
exponential expansion for the small true vacuum bubble. However, the quantum potential
rapidly trends to zero when the universe grows up, $Q(a)\sim -a^{-2-2|p-1|}$, which is too tiny
to support the accelerating expansion of the universe.

When $p=1$, the wavefunction of the vacuum universe is
\begin{equation}
\psi(a)=ic_{1}-c_{2}\ln a. \nonumber
\end{equation}
It is easy to obtain that the quantum potential
$$Q(a)= -\frac{c_1^2 c_2^2}{a^2 \left(c_2^2 \ln^2 a+c_1^2\right){}^2}.$$
The quantum potential approaches infinity when the universe
is very small $a\rightarrow 0$, but rapidly tends to zero when the universe grows up.

\subsection{Quantum potential of the universe with a scalar field}

In quantum cosmology, a widely used model to describe the universe is the minisuperspace
model containing a scalar field. In this model, the WDWE can be written as~\cite{av86,dlw00}
\begin{equation}
\left[  \frac{1}{a^{p}}\frac{\partial}{\partial a}a^{p}\frac{\partial
}{\partial a}-\frac{1}{a^{2}}\frac{\partial^{2}}{\partial\phi^{2}}%
-U(a,\phi)\right]  \psi(a,\phi)=0,  \label{WDWE:scalar}
\end{equation}
where the classical potential $U(a,\phi)=a^{2}[k-a^{2}V(\phi)]$, $V(\phi)$ is the potential of the scalar field,
and $\psi(a,\phi)$ is the wavefunction of the universe.
Inserting $\psi(a,\phi)=R(a,\phi)e^{iS(a,\phi)}$ into Eq.~(\ref{WDWE:scalar}) and
separating the equation into real and imaginary parts, one obtains two equations:
\begin{eqnarray}
S_{aa}+2\frac{R_{a}S_{a}}{R}+\frac{pS_{a}}{a}-a^{-2}\left(S_{\phi\phi}+\frac{2R_{\phi}S_{\phi}}{R}\right)  &  =&0,\label{wdwe:s1}\\
(S_{a})^{2}-a^{-2}{S_{\phi}}^2+U(a,\phi)+Q(a,\phi) &  =&0, \label{wdwe:s2}%
\end{eqnarray}
where $X_a$ denotes $X$ derivatives with respect to $a$, etc. The quantum potential $Q(a,\phi)$ is
\begin{equation}
Q(a,\phi)=-(\frac{R_{aa}}{R}+\frac{p}{a}\frac{R_{a}}{R})+\frac{R_{\phi\phi}}{a^2 R}. \label{quantum:potential:scalar}%
\end{equation}
The quantum Hamilton-Jacobi theory gives the guidance relations~\cite{npn12}
\begin{eqnarray}
\partial_{a}S(a,\phi)  &  =&-a\dot{a}, \label{grphia} \\
\partial_{\phi}S(a,\phi)  &  =&a^{3}\dot{\phi}.\label{grphiphi}
\end{eqnarray}
Inserting Eq.~(\ref{grphia}) and~(\ref{grphiphi}) into Eq.~(\ref{wdwe:s2}), we can obtain
\begin{equation}
  H^2=\dot{\phi}^2-\frac{U(a,\phi)+Q(a,\phi)}{a^4}. \label{Hubble:scalar}
\end{equation}

Generally, is it hard to obtain analytic solutions to Eq.~(\ref{WDWE:scalar}), and therefore,
we cannot obtain analytic solutions of the quantum potential and the quantum trajectory of the universe.
 Considering the slow changing of the current universe, if there is indeed a scalar field, it is likely
to be a slow-rolling scalar field. In fact, a slow-rolling scalar field itself may be considered as a cosmological
constant that can accelerate the expansion of the universe, in principle. However, we still want to know whether
the quantum potential of the universe is comparable to
the dark energy as required when a slow-rolling scalar field exists, particularly the total effects of the
classical potential of the scalar field and the quantum potential.

Since the scalar field is slow-rolling, the dependency on $\phi$ in Eq.(\ref{WDWE:scalar}) can be effectively ignored.
In this case, the WKB solutions of Eq. (\ref{WDWE:scalar}) are found to be~\cite{kw99}
\begin{equation}
\Psi (a,\phi )=\left\{\begin{aligned}\frac{A_{\pm } e^{\frac{\mp i}{4} \pi }}{a^{(p+1)/2}\left(a^2 V -1\right)^{1/4}}\exp \left[\frac{\pm i\left(a^2 V-1\right)^{\frac{3}{2}}}{3 V}\right],  \ \textit{for} \ a^2 V>1.\\ \nonumber
\frac{B_{\pm } }{a^{(p+1)/2}\left(1-a^2 V  \right)^{1/4}}\exp \left[\frac{\pm \left(a^2 V-1\right)^{\frac{3}{2}}}{3 V}\right],  \ \textit{for} \ a^2 V<1.  \nonumber
\end{aligned}
\right.
\end{equation}
The specific form of the wave function should be determined by the boundary conditions.
Here we focus on the case of present universe that satisfies the condition $a^2 V>1$.
Using the Hartle-Hawking boundary condition, the wave function can be determined as~\cite{kw99}
\begin{equation}
\Psi_{\mathrm{HH}}=\frac{C e^{\frac{1}{3V}}\cos\left[\frac{(a^2 V-1)^{3/2}} {3V}-\frac{\pi}{4} \right]}  {a^{(p+1)/2}\left(a^2 V -1\right)^{1/4}}.
\end{equation}
Since $\Psi_{{\rm HH}}$ is a pure real function, one can get that $S(a,\phi)=0$ and hence
the universe cannot accelerating expansion, i.e., $\dot{a}=0$ in Eq.~(\ref{grphia}).
One can also get the total effects of the classical potential and the quantum potential as $Q(a,\phi)+U(a,\phi)=0$
from Eq.~(\ref{wdwe:s2}).

Vilenkin's tunneling proposal suggests that the wave function restricted by $a^2 V>1$ should be~\cite{kw99}
\begin{equation}
\Psi_{{\rm V}}=\frac{e^{-\frac{1}{3V}}e^{i \pi/4}}  {a^{(p+1)/2}\left(a^2 V -1\right)^{1/4}} \exp{\frac{-i (a^2 V-1)^{3/2}}{3V}}.
\end{equation}
In this case, we can obtain that
\begin{eqnarray}
R(a,\phi)&=&\frac{e^{-\frac{1}{3V}}} {a^{(p+1)/2}\left(a^2 V -1\right)^{1/4}} ,\\
S(a,\phi)&=&- \frac{(a^2 V-1)^{3/2}}{3V}+\frac{\pi}{4}.
\end{eqnarray}
Using the guidance relation (\ref{grphia}), we obtain the present Hubble parameter
\begin{equation}
H_0^2=\left(\frac{\dot{a}}{a}\right)\approx V.
\end{equation}
This shows that the evolution of the grown-up universe is almost determined by the classical potential of the scalar field.
Inserting $R(a,\phi)$ into Eq.~(\ref{quantum:potential:scalar}), and taking a specific but widely used scalar field,
$V(\phi)=m^2 \phi^2 /2$, we obtain
\begin{equation}
Q(a)=\frac{(p-4)(p+2)}{4a^{2}}+\frac{1}{a^2}\left( \frac{2m^2}{9V^3}-\frac{4m^2}{3V^2}+\frac{3m^2}{8V}   \right).
\end{equation}
One can get that $Q(a,\phi)/ U(a,\phi)\sim a^{-6}$ for $a\gg 1$. This clearly shows that the quantum
effects of the universe are ignorable compared with classical potentials when the universe
is large. It is not the quantum potential but the slow-rolling scalar field play the role of dark energy to promote the
accelerating expansion of the universe in this case.

Another interesting case is the Wheeler-DeWitt equation containing a free massless scalar,
which has been studied in~\cite{ps03}. It was shown that the quantum effects are significant near the region $a=1$, and
quantum potential becomes negligible compared with classical potential when $a\gg 1$~\cite{ps03}.
This means that when the universe with a free massless scalar field  grows up, quantum effects have disappeared,
and the behavior of the universe is completely governed by classical potential, which is similar to our results above.

\section{Discussion and conclusion}\label{Discussion:and:conclusion}

In Eqs. (\ref{wave:function}) and (\ref{psi0}), we have obtained general solutions of the
wavefunctions of the universe. We want to discuss some special case of the wavefunctions, e.g.,
the wavefunction is purely real. In the case of $\psi(r)=R(r)$, it is easy to get $S(r)=0$, which means
there is no acceleration for the universe.
In fact, according to Schr\"odinger equation, one can obtain that
\begin{equation}
\nabla^2 \left[-\frac{\hbar^2}{2m^2} \left(\frac{1}{R}\nabla^2R \right)+V\right]\equiv0.
\end{equation}
Therefore, a quantum potential obtained from a purely real, (or purely imaginary, i.e., $R(r)=0$) wavefunction
cannot provide an accelerating universe~\footnote{In \cite{as15}, they used a real wavefunction to explain
the quantum potential (i.e., the quantum correction term) as current cosmology constant. It is obvious that
a real wavefunction cannot provide an accelerating universe, neither accelerating expansion nor accelerating shrink.}.

In summary, based on WDWE, we have derived the quantum modified Friedmann equations by using de Broglie-Bohm
quantum trajectory approach. We have shown that quantum effects of the universe cannot play a role as dark energy for a large
universe, neither for a matter-dominated universe, nor for a vacuum universe. For a universe with a slow-rolling
scalar-field, the behavior of the large universe depends on the boundary condition of wavefunction.
The Hartle-Hawking boundary condition cannot provide accelerating expansion since the wavefunction is real,
and the effect of the classical scalar field is completely offset by the quantum potential. On the other side,
Vilenkin's tunneling wavefunction allows a large universe with a slow-rolling scalar field to accelerate expansion.
Our results indicate that we should reexamine more carefully whether our universe is expanding at an accelerating rate or not.
Since the quantum effects of a vacuum universe or a matter-dominated universe cannot provide accelerated expansion,
we should seriously consider some other theoretical models,
such as the slow-rolling scalar field model, the field particle model~\footnote{According to the
current value of dark energy density, it has recently been proposed that there may be a field particle with mass $m\sim2\times10^{-33}eV$,
whose quantum potential can be explained as cosmological constant~\cite{grl19}. If such particles can be observed in the future experiments,
then the mystery of dark energy can be uncovered.}, or the modified gravity theory~\cite{cfp12}.

\textit{Acknowledgement}.---This work was supported by the National Natural Science Foundation of
China (Grant No. 11725524), Shaanxi Youth Outstanding Talent Support Plan and Shaanxi
Natural Science Foundation(Grant No.2019JQ-900) and the Fundamental Research Funds of Xianyang Normal
University (Grant No. XSYGG201802).

\end{document}